\documentclass[aps,pra,twocolumn,showpacs]{revtex4}

\usepackage{graphicx}
\usepackage{amsmath,amssymb,latexsym}

\newcommand{\setZ}{\mathbb{Z}}
\newcommand{\xor}{\overline{\lor}}
\newcommand{\XOR}{\textsc{xor}}
\newcommand{\NOPA}{\text{NOPA}}
\newcommand{\EPR}{{EPR}}
\newcommand{\CHSH}{{CHSH}}

\begin{document}
\title{Bell Inequalities for Position Measurements}

\author{Jan-\AA ke Larsson}\affiliation{Matematiska Institutionen,
  Link\"opings Universitet, SE-581 83 Link\"oping, Sweden}
\email{jalar@mai.liu.se}

\date{October 22, 2003}

\begin{abstract}
  Bell inequalities for position measurements are derived using the
  bits of the binary expansion of position-measurement results.
  Violations of these inequalities are obtained from the output state
  of the Non-degenerate Optical Parametric Amplifier.
\end{abstract}

\maketitle


In the Einstein-Podolsky-Rosen (EPR) paradox devised in Ref.\ 
\cite{EPR}, a main ingredient is position measurements, and it has
been a long-standing controversy whether such measurements together
with momentum measurements provide non-local statistics or not. While
it certainly is true that the original EPR paper only is intended to
ask the question of completeness, many have put considerable thought
into the of the possible non-locality of this system.  It would seem
that this question was put to rest by Bell in Ref.\ \cite{Bell86}
where he presented a local realist model for position and momentum
measurements on the original \EPR\ state, constructed using the Wigner
function representation of the \EPR\ state as a joint probability of
the measurement results. The Wigner function generally has all the
properties of a probability measure except one: it can be negative (a
proper probability measure is always positive). However, in this
particular case the Wigner function is positive, so it can be used as
a proper probability measure.

One could be led to think that this implies that the \EPR\ state can
be described by a local realist model, but this is not the case: the
important thing to note is the statement \emph{``for position and
  momentum measurements''}.  That is, nothing is said about
\emph{other} measurements; the quantum state contains more than just
information about position and/or momentum. In fact, if one instead
uses measurements of \emph{parity}, one can interpret the Wigner
function as a correlation function for these parity measurements, and
then certain regularized \EPR\ states \emph{are} nonlocal, e.g., the
Non-degenerate Optical Parametric Amplifier (NOPA) state
\cite{BanaWodk}. The next step was taken by Chen \emph{et al}
\cite{CPHZ} who used parity of number to violate a Bell inequality.
Finally, in Ref.~\cite{Jalar03a} it was shown that the full number
operator together with suitable other operators can be used to violate
the Bell inequality. The aim of this paper is to show that the
position operator itself together with suitable other operators also
can be used to violate the Bell inequality, in a sense, deriving a
Bell inequality more suited to the original EPR setting.

In the standard Bell inequality there is a bipartite spin-$\tfrac12$
system (see e.g., Ref.~\cite{Bell64,CHSH}); at each site a
spin-$\tfrac12$ measurement is made, and the direction along which the
measurement is made constitute a local parameter. In general, this
parameter is a direction in space, but in this paper for simplicity
only one angular parameter will be retained, in quantum notation,
\begin{equation}
  s_{\phi}=\cos\phi\, s_z+\sin\phi\,s_x
  \label{eq:1}
\end{equation}
The shorthand notation $s_{\alpha}s'_{\gamma}$ will be used to denote
the quantum operator $s_{\alpha}\otimes s_{\gamma}$ below. The
\emph{results} of the individual measurements $s_{\alpha}\otimes{}I$
and $I \otimes s_{\gamma}$ will be denoted $S_{\alpha}$ and
$S'_{\gamma}$. That is, these are the classical $\pm1$
(``up''/``down'') values registered from each local measurement, e.g.,
written down on a piece of paper or similar.  The question is now if
these results can be described under the assumption of local realism:
\begin{enumerate}
\renewcommand{\theenumi}{\roman{enumi}}
\renewcommand{\labelenumi}{(\theenumi)}
\item \label{p:Realism} \emph{Realism:} There is a classical
  probabilistic model where the results depend on a ``hidden
  variable'' $\lambda$, i.e.,
  \begin{equation}
    \begin{split}
      S_{\alpha}&=S_{\alpha}(\lambda),\\
      S'_{\gamma}&=S'_{\gamma}(\lambda).
    \end{split} \label{eq:2}
  \end{equation}
\item \label{p:Locality} \emph{Locality:} The model is local, such
  that measurement settings at one site does not affect the other
  site, i.e.,
  \begin{equation}
    \begin{split}
      S_{\alpha}(\lambda)&\text{ is independent of }\gamma,\\
      S'_{\gamma}(\lambda)&\text{ is independent of }\alpha.
    \end{split}
  \end{equation}
\item \label{p:Result} \emph{Result restriction.} The measurement
  results are restricted in size:
  \begin{equation}
    |S_{\alpha}(\lambda)|\le 1,\quad |S'_{\gamma}(\lambda)|\le 1.
  \end{equation}
\end{enumerate}
When this is the case, we have the \CHSH\ inequality \cite{CHSH}
\begin{equation}
  \begin{split}
    &\big|E(S_{\alpha}S'_{\gamma})
    +E(S_{\alpha}S'_{\delta})\big|
    +\big|E(S_{\beta}S'_{\gamma})
    -E(S_{\beta}S'_{\delta})\big| \leq2.
\end{split}
\label{eq:Bell}
\end{equation}
This inequality is violated by the entangled state
\begin{equation}
  \label{eq:3}
  |\psi\rangle
  =\tfrac{1}{\sqrt2}
  \big(|+_z+_z\rangle+|-_z-_z\rangle\big),
\end{equation}
at the settings $\alpha=0$, $\beta=\frac\pi2$, $\gamma=\frac\pi4$, and
$\delta=-\frac\pi4$, for which the corresponding \emph{quantum}
expression
\begin{equation}
  \big|\langle s_{0}s'_{\frac\pi4}\rangle
  +\langle s_{0}s'_{-\frac\pi4}\rangle\big|
  +\big|\langle s_{\frac\pi2}s'_{\frac\pi4}\rangle -\langle
  s_{\frac\pi2}s'_{-\frac\pi4}\rangle\big|
  =2\sqrt2>2.
  \label{eq:4}
\end{equation}

\begin{figure}[htbp]
  \begin{center}
    \includegraphics[width=\linewidth]{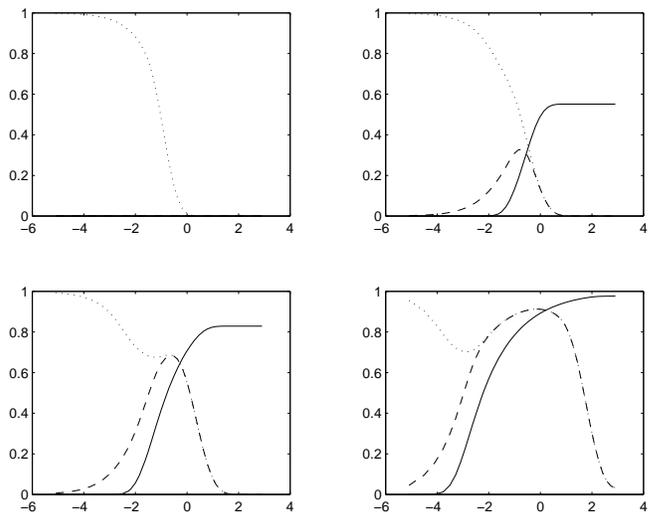}
    \caption{Calculation of $\langle s_{z,l}s_{z,l}'\rangle$ 
      (solid line), $\langle s_{x,l}s'_{x,l}\rangle$ (dotted line),
      and $\langle s_{y,l}s'_{y,l}\rangle$ (dashed line) for %
      $0.03\le l \le 7.5$ (the horizontal scale is logarithmic, i.e.,
      $\log_2 l$) and $r=0$, $0.5$, $1$, and $2$ from top left to
      bottom right.}
    \label{fig:1}
  \end{center}
\end{figure}

The idea now is to extend this to position measurements.  To do this,
we choose an interval length $l$, quite arbitrarily for now, and
denote the position operator $q$, since the letter $x$ is used for
other purposes. Now, let us determine whether or not the measured
value of the position $Q$ is in a box of length $l$, at position $nl$
from the origin for integer $n$. This corresponds to an operator
$k_{n,l}$ that projects a state $\psi$ (in position representation)
onto the subspace of states with support only in $nl\le q < nl+l$,
\begin{equation}
  k_{n,l}\psi(q)=
  \begin{cases}
    \psi(q),& nl\le q < nl+l\\
    0, & \text{otherwise.}
  \end{cases}
\end{equation}
We can now write down an operator that will become the equivalent of a
spin-$z$ operator,
\begin{equation}
  s_{z,l}=\sum_{n=-\infty}^{\infty}(-1)^n k_{n,l}.
  \label{eq:5}
\end{equation}
So far, this operator only has the same eigenvalues as the usual
spin-$z$ operator ($\pm1$), but the construction can be extended to a
full pseudo-spin system. To do this we need spin-step operators, and
these can be constructed from a box-translation operator that
translates a state $\psi$ by $l$ length units and then projects into
the same subspace as used above,
\begin{align}
  t_{n,l}\psi(q) &=
  \begin{cases}
    \psi(q+l),& nl\le q < nl+l\\
    0,                      & \text{otherwise,}
  \end{cases}
\intertext{with adjoint}
  t^{\dag}_{n,l}\psi(q) &=
  \begin{cases}
    \psi(q-l),& nl+l\le q < nl+2l\\
    0,                      & \text{otherwise.}
  \end{cases}
\end{align}
The positive spin-step should take an eigenstate of $s_{z,l}$ with
eigenvalue $-1$ to one with eigenvalue $+1$, and its adjoint should do
the opposite:
\begin{equation}
    s_{+,l}=(s_{-,l})^{\dag} 
    =\sum_{n=-\infty}^{\infty}t_{2n,l}.
\end{equation}
Via the usual
\begin{equation}
    s_{x,l}=s_{+,l} + s_{-,l},\qquad s_{y,l} = -i (s_{+,l} - s_{-,l})
\end{equation}
we have constructed a pseudo-spin system,
\begin{equation}
    [s_{x,l}, s_{y,l}] = 2i s_{z,l},\quad \ldots,\quad \text{and}\quad 
    s_{x,l}^2= s_{y,l}^2 =  s_{z,l}^2 = \mathbb{I}.
\end{equation}
We can now derive a violation of the simple Bell inequality
(\ref{eq:Bell}) above from our two-mode Gaussian \NOPA
state. In position expansion,
\begin{equation}
  \left|\psi_{\NOPA}\right\rangle=
  \frac{1}{\sqrt{\pi}}e^{qq'\sinh 2r-\frac12 
    \big(q^2+(q')^2\big)\cosh 2r}.
  \label{eq:6}
\end{equation}
By symmetry, measurements along orthogonal directions are uncorrelated
at the two sites, e.g., %
$\langle s_{z,l}s_{x,l}'\rangle=0$. The correlations for equal
directions must unfortunately be calculated by numeric integration of
a two-dimensional Gaussian on $l\times l$ squares. For varying
squeezing $r$ and minimal length $l$, we have data as shown in
Fig.~\ref{fig:1}.

In Fig.~\ref{fig:1} there are some interesting features. First, there
is a high correlation in the $s_{x,l}$ and $s_{x,l}'$ results when $l$
is small. This is perfectly natural from the construction of the
operator; it is such that if the wave-function is smooth
\footnote{Mathematically speaking, it suffices that the wave-function
  is continuous, bounded, and has a bounded derivative.}, then
\begin{equation}
  \label{eq:7}
  \lim_{l\rightarrow\infty}
  \langle s_{x,l}\rangle=1.
\end{equation}
We have the completely classical property that measurement of
$s_{x,l}$ always yields the result 1 at very small $l$. When the
squeezing is increased, the slope of the wave function is larger, so
that one needs to go to a smaller $l$ to get this effect, but at a
given squeezing, there is always an $l$ at which this sets in.

Furthermore, there is a high correlation between $s_{z,l}$ and
$s_{z,l}'$ when $l$ is large, but for a different reason; note that
$\langle s_{z,l}\rangle=0$ for all $l$. When $l$ is large, almost all
of the probability falls into the four $l\times l$ squares around the
origin. When the wave function is squeezed, more probability falls
into the two boxes where the $s_z$ and $s_z'$ results are the same.
For fixed squeezing $r$, the limit for large $l$ can be calculated
explicitly as
\begin{equation}
  \label{eq:8}
  \lim_{l\rightarrow\infty}
  \langle s_{z,l}s_{z,l}'\rangle =\frac{2}{\pi} 
  \arctan\big(\sinh (2r)\big).
\end{equation}
This value corresponds well with the values seen in Fig.~\ref{fig:1}.

The interesting regime is where all three correlations %
$\langle s_{x,l}s_{x,l}'\rangle$, $\langle s_{y,l}s_{y,l}'\rangle$,
and $\langle s_{z,l}s_{z,l}'\rangle$ are high, remember that
the state in Eq.~\eqref{eq:3} has %
$\langle s_{x}s_{x}'\rangle=\langle s_{y}s_{y}'\rangle= \langle
s_{z}s_{z}'\rangle=1$, and this is the reason for the high violation
of the Bell inequality. The correlations are somewhat lower here, so
consequently maximal violation is not to be expected. Of course, high
squeezing is needed so that the three correlations are high in the
intermediate regime.

\begin{figure}[htbp]
  \begin{center}
    \includegraphics[width=\linewidth]{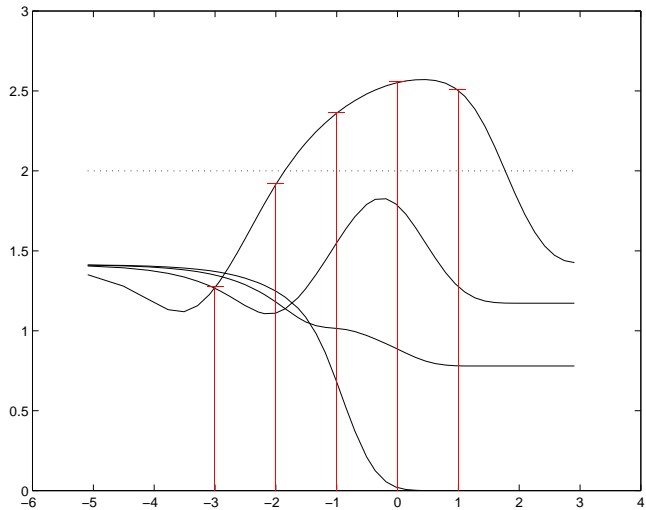}
    \caption{The expression~(\ref{eq:4}) for lengths
      $0.03\le l \le 7.5$ and $r=0$, $0.5$, $1$, and $2$ from the
      lowest to the highest curve. The Bell inequality is violated
      whenever the value exceeds 2, e.g., when $l=1$ and $r=2$. The
      indicated values for $r=2$ will be used below.}
    \label{fig:2}
  \end{center}
\end{figure}

To calculate the interesting quantum expression~\eqref{eq:4} for
different $l$s, we bring in pseudo-angles $\alpha=0$,
$\beta=\frac\pi2$, $\gamma=\frac\pi4$, and $\delta=-\frac\pi4$ as in
Eqs.~(\ref{eq:1}) and ~(\ref{eq:4}). The standard values of the
pseudo-directions are used here, although these probably do not give
the maximal violation. Nonetheless, in Fig.~\ref{fig:2} there is a
clear violation of the Bell inequality in the intermediate region for
$l$, when the squeezing $r$ is high enough.



Returning to position, write the measured position in base 2, e.g.,
\begin{equation}
  \label{eq:9}
  Q=5.296875_{10}=101.0100110_2,
\end{equation}
which is an expression on the form
\begin{equation}
  \label{eq:10}
  Q=\sum_{k=-\infty}^{\infty}2^k B_k.
\end{equation}
Note that when the ``zeroth'' bit $B_0$ is 1, the corresponding
measured value $S_{z,1}$ (i.e., $l=1$) would have been $-1$. Also when
the zeroth bit $B_0$ is 0, the corresponding measured value $S_{z,1}$
would have been $+1$. We have $S_{z,1}=1-2B_0$, and in general
\begin{equation}
  \label{eq:11}
  S_{z,2^k}=1-2B_k.
\end{equation}
Multiplying two spin values corresponds to taking \XOR{} on the
appropriate bit values (see Ref.~\cite{Jalar03a}), so that
  \begin{equation}
    \label{eq:12}
    SS'=1-2B\xor B'
  \end{equation}
It is also simple to introduce a parameter $\phi$ so that 
\begin{equation}
  \label{eq:13}
  S_{\phi,2^k}=1-2B_{\phi,k}
\end{equation}
Using this last correspondence and Eq.~(\ref{eq:12}), the Bell
inequality (\ref{eq:Bell}) becomes
\begin{equation}
  \begin{split}
    &\big|E(B_{\alpha,k}\xor B'_{\gamma,k})
    +E(B_{\alpha,k}\xor B'_{\delta,k})-1\big|
    \\&\quad
    +\big|E(B_{\beta,k}\xor B'_{\gamma,k})
    -E(B_{\beta,k}\xor B'_{\delta,k})\big| \leq1.
  \end{split}
  \label{eq:BellB}
\end{equation}
The right-hand side is a factor of two lower than before, but the
sensitivity to noise and such-like is the same as before, we have just
changed notation.

In our new notation, we can use the corresponding quantum
expression to obtain a quantum bit (``qubit'')
\begin{equation}
  s_{\phi,2^k}=1-2b_{\phi,k}
\end{equation}
and a ``quantum \XOR''
\begin{equation}
  ss'=1-2b\xor b'
\end{equation}
which is a noncommutative operation when used on the same subsystem
(see Ref.~\cite{Jalar03a}). We obtain a quantum expression
corresponding to~\eqref{eq:4},
\begin{equation}
  \begin{split}
    &\big|\langle b_{0,k}\xor b'_{\frac\pi4,k}\rangle +\langle
    b_{0,k}\xor b'_{-\frac\pi4,k}\rangle-1\big| \\&\quad\quad
    +\big|\langle b_{\frac\pi2,k}\xor b'_{\frac\pi4,k}\rangle -\langle
    b_{\frac\pi2,k}\xor b'_{-\frac\pi4,k}\rangle\big|.
  \end{split}
  \label{eq:14}
\end{equation}
In Fig.~\ref{fig:2} (vertically scaled by the factor $\frac12$) there
is a simultaneous, separate violation for $r=2$, and $k=1$, $0$, and
$-1$, while there is no violation for, e.g., $k=-2$ and $-3$.

We have used our continuous hierarchy of pseudo-spin systems indexed by
the (real-valued) variable $l$, but only for values $l=2^k$. The
reason is that, to talk about \emph{simultaneous} violation, the
pseudo-spin systems must commute with each other. The spin-operator
$s_{z,l}$ trivially commutes with the spin operator $s_{z,cl}$ for any
$c$, but it does not commute with $s_{x,cl}$ for all $c$s, for example
when $c=1$. However, these two \emph{do} commute in an important
special case, when $c$ is an even number. This is because changes to
the wave function caused by $s_{z,l}$ are wholly within each interval
translated by, e.g., $s_{x,2ml}$, and the changes are the same within
each such interval. It is possible to show that the whole pseudo-spin
system for length $l$ commutes with that for length $2ml$. Thus, there
is an infinite commuting hierarchy of spin systems,
\begin{equation}
  \{s_{z,2^kl},s_{+,2^kl}=(s_{-,2^kl})^{\dag}\}_{k\in\setZ}.
\end{equation}

The measured position $Q$ was expanded in base 2.  The same can be
done for the quantum operator
\begin{equation}
  \label{eq:15}
  q=\sum_{k=-\infty}^{\infty}2^k b_k=
  \sum_{k=-\infty}^{\infty}2^k \frac{1-s_{z,2^k}}2.
\end{equation}
Here it becomes really important that $s_{z,2^k}$ all commute.  Since
also $s_{x,2^k}$ all commute, and $s_{y,2^k}$ all commute for
different $k$, it is meaningful to write, e.g.,
  \begin{equation}
    \label{eq:16}
    q_x=\sum_{k=-\infty}^{\infty}2^k b_{x,k}=
    \sum_{k=-\infty}^{\infty}2^k \frac{1-s_{x,2^k}}2,
  \end{equation}
and similarly for $q_y$. These would be pseudo-positions obtained by
qubit rotations of the original position operator $q$.  To emphasize
the presence of a pseudo-spin system in the ordinary position
measurement, one could write $q_z=q$, but this will not be done below.
The infinite expansions have their own, more mathematical problems
which will not be discussed further here, note that violations only
occur for finite $k$ for our continuous, localized
$|\psi_{\NOPA}\rangle$ state. Let us use the truncated expression
\begin{equation}
  \label{eq:17}
  Q_{\phi,\text{\textvisiblespace}}=2B_{\phi_1,1}+
  B_{\phi_0,0}+
  \frac{B_{\phi_{-1},-1}}2+
  \frac{B_{\phi_{-2},-2}}4+
  \frac{B_{\phi_{-3},-3}}8.
\end{equation}
By using ineq.~(\ref{eq:BellB}) for each bit in the above expression,
and letting $\xor$ denote \emph{bitwise} \XOR, we obtain
($2+1+\frac12+\frac14+\frac18=3.875$)
\begin{equation}
  \begin{split}
    &\big|E(Q_{\alpha,\text{\textvisiblespace}}\xor
    Q'_{\gamma,\text{\textvisiblespace}})
    +E(Q_{\alpha,\text{\textvisiblespace}}\xor
    Q'_{\delta,\text{\textvisiblespace}})
    -3.875 \big|\\
    &\quad+\big|E(Q_{\beta,\text{\textvisiblespace}}\xor
    Q'_{\gamma,\text{\textvisiblespace}})
    -E(Q_{\beta,\text{\textvisiblespace}}\xor
    Q'_{\delta,\text{\textvisiblespace}})\big|
    \leq 3.875.
  \end{split}
  \label{eq:Bell2}
\end{equation}
Reading off the quantum values in Fig.~\ref{fig:2}, we get
\begin{equation}
  \begin{split}
    &\big|\langle q_{\text{\textvisiblespace}}\xor
    q'_{\frac\pi4,\text{\textvisiblespace}}\rangle +\langle
    q_{\text{\textvisiblespace}}\xor
    q'_{-\frac\pi4,\text{\textvisiblespace}}\rangle -3.875 \big|\\
    &\quad\quad
    +\big|\langle q_{\frac\pi2,\text{\textvisiblespace}}\xor
    q'_{\frac\pi4,\text{\textvisiblespace}}\rangle -\langle
    q_{\frac\pi2,\text{\textvisiblespace}}\xor
    q'_{-\frac\pi4,\text{\textvisiblespace}}\rangle\big| \\
    &\quad\approx 2\times1.25+1.28+\frac12\times1.18
    \\&\quad\quad+\frac14\times0.96+\frac18\times0.64\approx4.7
    >3.875.
  \end{split}
  \label{eq:Bell4}
\end{equation}
Here, the operator $q_{\text{\textvisiblespace}}$ is the usual
position operator $q$, truncated to the bits in question. The
truncation is done for several reasons. First and perhaps most
important, it is not beneficial to include bits where there is no Bell
violation, but it is possible to include such bits as seen above.
Second, we would expect the measurement devices to be of finite size,
so that there is a natural upper bound; note that finite size devices
do not introduce any substantial error in the above reasoning, since
the Gaussian is well localized.

Finally, the devices have finite resolution so there is a natural
lower bound. On should perhaps remark that this is a question of
resolution rather than accuracy. Low accuracy would yield bit-result 0
when there should have been a bit-result 1 and vice versa; there would
have been more noise in the data. However, realization of the
measurements corresponding to the operators defined here is a
difficult issue. At present there is no implementation known to the
author that would give the desired properties. Note that the qubits
are in principle present in any position measurement; consequently
they could be used in quantum information processing. This latter
possibility in itself will perhaps provide a motive to realize this
proposal, rather than the Bell violation. In any case, a full analysis
of possible experimental errors is better postponed until an
experimental procedure is known.

We have a Bell inequality where one of the operators correspond to
position, and a violation from the NOPA state. When the squeezing
increases, the violation of the inequality increases, and it is to be
expected that infinite squeezing will give a maximal violation at each
qubit of the position operator. This, in turn, means that the original
EPR state (infinite squeezing) will violate the above inequalities
maximally. Because of lack of space and the more mathematical nature
of a proper statement of this, it will be deferred to a later
publication.

One possible extension is to use qutrits (spin-1 correspondence)
instead of qubits in the approach, or indeed so-called qu$N$its
(spin-$(N-1)/2$ correspondence) for arbitrary $N$, together with an
inequality more suited to such a situation \cite{KGZMZ,CGLMP,CWKO}.
There is also the question of what settings are the best to put in
place of $\alpha$, $\beta$, $\gamma$, and $\delta$ at finite
squeezing. The violation is not at its highest when using the settings
from this paper, as this paper concentrates on the more conceptual
issues. The maximum is probably best found by numerical optimization,
given that we have numerical data only.

We have achieved what we set out to do. And the perhaps unsurprising
conclusion is that the NOPA state cannot be described by a local
realist model, despite having a strictly positive Wigner function.
Note, however, that it was necessary to augment $q$ with operators not
directly related to the momentum operator $p$. Nevertheless, this
formulation is somewhat closer to the original EPR paper than the
traditional Bell approach.

\begin{acknowledgments}
  This work has been supported by the Swedish Science Council.
\end{acknowledgments}



\end{document}